\documentclass[onecolumn]{article}

\usepackage[dvips]{graphicx}
\usepackage{amssymb,amsfonts,amsmath}
\usepackage{amsmath,epsfig}

\date{}

\begin{document}

\title{Homologous Control of Protein Signaling Networks.}

\author{D. Napoletani \thanks{ Corresponding author address:
Center for Applied Proteomics and Molecular Medicine,
George Mason University, Manassas, VA 20110 USA. email:
dnapolet@gmu.edu.}, M.
Signore\thanks{Department of Hematology, Oncology and Molecular
Medicine, Biotechnology Division, Istituto Superiore di
Sanit\`{a}, Viale Regina Elena 299, 00166, Rome, Italy}, T.
Sauer\thanks{Department of Mathematical Sciences, George Mason
University, Fairfax, VA 22030 USA}, L. Liotta
\thanks{Center for Applied Proteomics and Molecular Medicine,
George Mason University, Manassas, VA 20110 USA.}, E.
Petricoin$^{\S}$}

\maketitle

\begin{abstract}

In a previous paper we introduced a method called augmented sparse
reconstruction (ASR) that identifies links among nodes of ordinary
differential equation networks, given a small set of observed
trajectories with various initial conditions. The main purpose of
that technique was to reconstruct intracellular protein signaling
networks.

In this paper we show that a recursive augmented sparse
reconstruction generates artificial networks that are homologous
to a large, reference network, in the sense that kinase inhibition
of several reactions in the network alters the trajectories of a
sizable number of proteins in comparable ways for reference and
reconstructed networks. We show this result using a
large in-silico model of the epidermal growth factor receptor
(EGF-R) driven signaling cascade to generate the data used in the
reconstruction algorithm.

The most significant consequence of this observed homology is that
a nearly optimal combinatorial dosage of kinase inhibitors can be
inferred, for many nodes, from the reconstructed network, a result potentially useful for a variety of applications in personalized medicine.

\end{abstract}

\section{Introduction}

One of the most intriguing and promising fields in medical
research is based on the assumption that the great amount of
information generated by high throughput technologies would allow
us to understand cancer's complexity at various levels. In recent
years, the completion of the Human Genome Project and other rapid
advances in genomics have led to increasing anticipation of an era
of genomic and personalized medicine, in which an individual's
health is optimized through the use of all available patient data,
including data on the individual's genome and its downstream
products.

Because variations in individuals' genetic profiles oftentimes
correlate with differences in how individuals develop diseases and
respond to treatment, personalized medicine supported by genetic
and genomic assays has the potential to facilitate optimal risk
identification, disease screening, disease diagnosis, therapy, and
monitoring \cite{1}, \cite{2}, \cite{3}. In addition to genomic
assays, proteomic and metabolomic signatures hold great potential
for serving as pillars of personalized medicine in the future
\cite{4}, \cite{5}, \cite{6}, \cite{7}, \cite{8}.

While personalized medicine guided by genomics is still in early
stages of development, individuals' genetic profiles are already
starting to be used to guide patient care. As some examples,
clinicians can obtain gene expression profiles of breast cancer
samples to guide management \cite{9}, genotypes of HIV samples to
identify the optimal antiretroviral regimen \cite{10}, and genetic
profiles of patients' cytochrome P450 drug metabolizing system to
guide the selection and dosing of pharmacotherapies \cite{11}.
However, it is the proteins that form the actual cell signaling and
metabolic networks within the cell. Indeed, for the new classes of
molecular targeted inhibitors, it is the proteins that are the
drug targets, not the genes, and the molecular networks are
underpinned by protein and protein phosphorylation.

Personalized medicine could be directed towards the generation of
protein-based molecular maps of cancer networks in order to target
malignant cells in their specific and unique context. The
usefulness of patient-tailored therapy comes from the potential
ability to depict patient-specific molecular circuitries and hence
translate each targeted treatment in a favorable clinical response
\cite{12}.

On the other hand, we still lack the ability to dynamically
measure and collect enough data from every protein/node within
networks with current methodologies. This restriction forces on us
a shift in mind set, in the sense that, rather than attempting a
full reconstruction and understanding of cell pathways, we should
search for equivalent, indistinguishable, classes of models that
project to the same network structure, in the sense that such
classes should ideally give rise, for each protein/node, to
trajectories that are qualitatively similar even when the details
of the topology of the connections among nodes differ.

The potential therapeutic implications of such an approach are evident if we consider the great heterogeneity of cancer patients. Individuals with similar stages of the disease show diverse therapeutic responses that are oftentimes not predictable on the basis of genetic mutation analysis. Primary cancer stem cells (CSCs) from a variety of tumors have been isolated and characterized and clear evidences exist of their association with tumor's resistance to chemo- and radio-therapy, making CSCs a useful and effective target for cancer research (\cite{12b}, \cite{12c}). Moreover, an extensive proteomic, genetic and drug sensitivity profiling of the NCI60 cell line set has recently highlighted the importance of direct studies on cancer cells in order to associate drug responsiveness to specific molecular signatures (\cite{12d}). Such studies, and clinical experience, call for a multi-targeted approach to cancer treatment and underscore the importance of molecular biomarkers (\cite{12e}). Therefore, an {\it in-silico} system capable of reconstructing the behavior of deregulated cancer cell signaling networks holds a significant experimental value in the prediction of therapeutic responses based on individual patients molecular characteristics.

Our approach to the problem of controlling protein signaling
networks starts with this broad methodological assumption, but it
necessarily moves further than that, since it is not yet clear
what we should consider as a measure of similarity of trajectories
for general, large networks. It is likely that properties such as
the shape of the trajectories, the location of their maxima, as
well as the value of the maxima themselves, are among the critical
factors in deciding whether or not a given signaling response is
triggered.

For example, activation of EGF-R, which is an upstream node and it
is governed only by the kinetics and thermodynamics of
EGF/EGF-receptor interaction and the biochemistry of the kinase
domain, is expected to be similar across cell lines. In contrast,
signals further down the cascade are modulated by many upstream
proteins, many of whose concentrations and rate constants impact
on the overall output.

Despite this complex network behavior, there is a strong
correlation between specific cell functions and the maximum
concentration of key proteins known to be involved in cell growth,
proliferation, survival and death, suggesting that
suppression/enhancement of the activity of specific nodes can be
seen potentially as a way to achieve the final goal of
disruptively interfering with the functioning of cancer cells.

The trajectories for each node are usually generated {\it in
vitro} by stimulation of cell lines and subsequent relaxation to
steady state, so that the extent of suppression of a node activity
can be determined by looking at the maximum value of a relatively
simple curve.

Even though the maximum activity of nodes is only one of many key
features of signaling networks, its accurate modification is by no
means an easy task. Agents directed at an individual target in the
network frequently show limited efficacy, poor safety and
resistance profiles, which are often due to factors such as
network robustness, redundancy, crosstalk, compensatory and
neutralizing actions and anti-target and counter-target
activities.

The ability to predict {\it in silico} the sensitivity of cancer
cells to the inhibition of multiple reactions would allow us to
combine drugs in order to achieve synergy and/or potentiation of
several orders of magnitude, while avoiding undesired effects on
normal cells. Systems-oriented approaches has already yielded
several clinical successes and drug-discovery efforts are now
focused towards near optimal combinatorial treatments that target
cell pathways at several sites.

Because the fundamental goal of a combinatorial approach to cancer
therapy is the control of the activity of specific nodes in the
network, we use it to define an operative notion of homologous
networks. We select a target node $N$, and a set of reactions $P$,
and we assume the following definition of {\bf homology of
networks:} {\it two networks are homologous (with respect to $N$
and $P$) if the activity of node $N$ reacts in a similar way to
the suppression of the given reactions $P$ performed by known
kinases}.

Note that this comparison can be made on very long time scales,
ideally on time intervals where the networks have each relaxed to
the steady state, so that the comparison of the networks can be
considered global.

{\bf Remark 1:} In Section 3 we define more formally similarity as
the concordance of the relative magnitude of the maximum
difference of the trajectories of the node $N$, starting from
equal initial conditions, when control of the reactions in the
networks, via kinase inhibition, is on, and when control is off.
In this way, we have a simple, even though partial, way to
determine how close two networks will react, for specific nodes,
to similar control schemes.

In a previous paper \cite {jtb} we introduced a method called
augmented sparse reconstruction (ASR) that identifies links among
nodes of ordinary differential equation (ODE) networks, given a
small set of observed trajectories with various initial
conditions. The main purpose of that technique was to reconstruct
intracellular protein signaling networks under the assumption that
most nodes interact with only a small fraction of the total number
of nodes in the network. We say in that case that the network is
{\it sparse} and such information  can greatly help in
reconstructing the network itself.

In this paper we show that augmented sparse reconstruction
generates artificial networks that are homologous to the initial
network, in the sense that kinase inhibition of several reactions
in the network alters the trajectories of a sizable number of
proteins in comparable ways. We show this surprising result using
an {\it in silico} model of the epidermal growth factor receptor
(EGF-R) driven signaling cascade to generate the initial observed
trajectories.

The most significant consequence of this observed homology is that
the the optimal combinatorial dosage of kinase inhibitors can be
inferred in many cases from the reconstructed network\footnote{A
patent application has been filed for the methods described in
this paper. Patent application number 12/959,096,
filed on 12/02/2010}. This result could be of great value for a
variety of applications in personalized medicine.

While there have been successful attempts to derive network models
from a limited number of perturbation experiments (see for example
the recent works \cite{new1}, \cite{new2}), we stress that our
method achieves a degree of reconstruction {\it and} dynamical
control for nodes of a network whose size far exceeds those tested
so far in the literature, with the exception of the very
interesting work in \cite{new3}, which uses sparsity in an
essential way, but that builds only a static model rather than a
dynamical one. A significant amount of information can be inferred
by static analysis, however a full network control in
non-stationary conditions can only be achieved in a dynamical
setting.

In section 2 we will show how the algorithm we described in
\cite{jtb} needs to be modified to take into consideration
knowledge of specific reactions that can be inhibited. Sections 3 and 4
are dedicated to comparison between an initial network and
partially homologous networks obtained from its trajectories by
augmented sparse reconstruction.

\section{Matching Pursuit for Augmented Sparse Reconstruction}

Because of our very limited understanding of the changes of
dynamics in large, perturbed networks (except in those cases when
the parameters of the model of individual nodes are only slightly
perturbed), it is daunting to set up an homologous network from
first principles, given a reference network.

We believe that the right approach to generate homologous networks
is to directly use the state space, in the sense that by modifying
or restricting information on the trajectories, we can use
reconstruction methods to give candidate homologous networks of a
reference network. Effectively, this is a signal processing
approach to network dynamics: optimal signal representation of the
trajectories becomes the main tool to explore network structure.

We select a well established model of the epidermal growth factor
receptor (EGF-R) signaling pathway as reference network
\cite{egfmod1}. The reason of such choice is the great importance
of the epidermal growth factor receptor signaling pathway in
cancer biology and the fact that it is one of the most
well-studied pathways that regulate growth, survival,
proliferation, and differentiation in mammalian cells \cite{14}.

In the EGF-R network, upon binding of the ligand, the receptors
dimerize and phosphorylate each other, thus generating docking
sites for five adaptor proteins and five enzymes. Signals from
ErbBs converge to molecules forming a bow-tie core and are
supposed to represent a versatile and conserved group of molecules
and interactions. The amplitude of EGF-R cascades reaches high
levels within minutes of stimulus and an important role is played
by the recycling mechanism of receptor molecules after signal
transduction, so that, in the absence of EGF molecules the system
relaxes back to steady state, in line with the generic description
of trajectories put forward in Section 1. The four human ErbB
receptors induce a wide variety of cellular responses thereby
generating a complex protein interaction network \cite{15}.

Due to its properties and involvement in tumor progression, the
EGF-R network inspired several experimental and mathematical
modeling studies \cite{16}, \cite{17}. Deregulation of EGF-R
signaling plays a key role in numerous cancers, including
glioblastomas, breast cancer, and non$-$small cell lung cancer
(NSCLC) \cite{18}.

Another reason to choose the EGF-R pathway as a reference network
is that, despite the fact that various agents have been developed
to target EGF-R, there is a need for improved strategies to
integrate anti-EGF-R agents with conventional therapies and to
explore combinations with other molecular targets \cite{19}.

In this work we use the differential equation model of EGF-R
network put forward in \cite{egfmod1}and \cite{egfmod2}. This
model assumes only linear and quadratic terms in the
representation of the derivative of the activity of each node of
the network. Linear terms correspond to uni-molecular interactions
and quadratic terms correspond to bimolecular interactions.

From the computational point of view, an important feature of this
network  is its large size (103 variables and 148 distinct
reactions). Most reconstruction techniques are not able to deal
with the reconstruction and control of very large networks, if the
experimental data are limited and noisy, and yet this is exactly
the size of networks that are of interest when exploring pathways
that may not be well understood.

In equation (1) we show the model of the network at a node $n$, in
the specific integral form that is used in augmented sparse
reconstruction; for a complete analysis of this integral model we
refer to \cite{jtb}. Essentially, equation (1) is nothing else
than the integral of a differential equation with linear and
quadratic terms, and with added random terms to make sure the
reconstruction algorithm is able to eliminate errors-in-variables
due especially to the presence of non-linear terms.
\begin{eqnarray}
x_n(t)-x_n(t_0)=a_{0n}+\sum_{i=1}^N l_{in} \int_{t_0}^t x_i
dt+\beta_q\sum_{i=1}^N\sum_{j=1}^N q_{ijn}\int_{t_0}^t x_ix_j dt
\\
\nonumber +\sum_{g=1}^G w_{gn} n_g.
\end{eqnarray}

Here the $\beta_q \leq 1$ represent positive attenuation
coefficients for the quadratic terms. The systems parameters at
node $n$ that we need to determine are: $a_{0n}$, $l_{in}$,
$i=1,...,N$, $q_{ijn}$, $i,j=1,...,N$. The $n_g$, $g=1,..,G$ are
discrete random vectors normally distributed, scaled to have norm
1 and multiplied by suitable parameters $w_{gn}$ to be determined
together with the system parameters.

The reconstruction algorithm of \cite{jtb} assumes sparsity of the
network, i.e. we assume that each node interacts with only a small
number of nodes compared with the total of possible nodes. This
assumption implies that the number of terms in each equation in
(1) with nonzero parameters is small compared to the total
possible number of terms.

Sparsity plays a crucial role in our method, since it allows to
use fast linear programming techniques in looking for the optimal
model that has as few terms as possible \cite{cds}, but just as
important for network reconstruction is the fact that our method
avoids a direct estimate of the derivative of the trajectories,
and that we augment the model with random terms. Despite these
adjustments, the quality of the reconstruction worsen for nodes
with many links, even when the total number of nonzero terms in
equation (1) is low compared to the total number of possible
terms.

Though sparsity methods are very powerful, when properly adapted
to networks, and they allow for significant inference of the
network under limited and noisy data, it is unlikely that
they, or any other currently known methods, will fully reconstruct
the network structure from very limited, coarse data.

Despite these limitations, the fundamental claim of this work is
that we can have homologous control despite our inability to gain
full reconstruction of the topology of a network. This claim is
intrinsically related, in ways that still need to be explored, to
a fundamental assumption of systems biology, i.e. the belief that
biological networks are robust under variations of the strength
and type of connections of the signaling pathways.

Robustness seems to be a consequence of several recurrent factors,
for example the bow-tie architecture (or hourglass structure) of
the EGF-R network is considered a characteristic feature for
robust evolvable systems \cite{kitano2004}. Another important
feature of robust biological networks is the fact that they show a
diverse array of molecules for input and output, that are
connected to the conserved core of the network with highly
redundant and extensively crosstalking pathways and feedback
control loops in various places in the pathway.

If the assumption of robustness is correct for most biological
networks, augmented sparse reconstruction may not recover the
exact network, but it may be sufficiently accurate to infer a
network that is homologous to the original one. We will see that
this possibility is realized for our EGF-R reference model.

In this work we assume that specific reactions must be present in
the reconstruction of the network, since we define homologous
systems with respect to the action of kinase inhibitors. In
\cite{jtb} there was no such constraint, therefore our main
objective in this section is to adapt the algorithm developed in
that work in such a way that it guarantees the presence of
specific reactions to be targeted with available kinase
inhibitors.

Signal processing sparsity methods, that are at the core of
augmented sparse reconstruction, are not able to guarantee the
presence of these individual reactions, since they are more
concerned with global optimality of the representation of each
node. We need therefore an adaptive, recursive augmented
reconstruction algorithm to extract the few terms in each equation
due to the chosen reactions, before we apply the augmented sparse
reconstruction algorithm to the whole representation system.

To understand the details of the reconstruction algorithm, we
first recall how individual reactions are put together in a
modular way to generate systems of differential equations
describing the network of \cite{egfmod1}, \cite{egfmod2}.

Suppose that phosphorylated proteins $x_i$ and $x_j$ are
interacting to phosphorylate protein $x_k$, and in the process
they get de-phosphorylated; this specific reaction can be modelled
\cite{egfmod1}, \cite{voit} as $v=ax_ix_j-bx_k$. Its effects on
the differential equations of the network are as follows: if we
let $\dot x_i=f_i(x_1,...,x_n)$, i.e. if we model the derivative
of the phosphorylated concentration of protein $x_i$ as a function
of the state of (possibly) all proteins, then, because the
reaction de-phosphorylate $x_i$, then $\dot
x_i=f_i(x_1,...,x_n)-v$, and similarly $\dot
x_j=f_j(x_1,...,x_n)-v$. On the contrary, since the specific
reaction increases the phosphorylation of $x_k$, we will have
$\dot x_k=f_k(x_1,...,x_n)+v$.

The immediate consequence of this modeling assumption is that if
we know that a simple quadratic reaction $v=ax_ix_j-bx_k$ is
involved in a network, then we know that the representation of the
derivative of $x_i$,$x_j$,$x_k$ will have a specific quadratic and
a specific linear term in the representation in (1). Only the {\it
parameters} of these terms will be unknown.

The more reactions we make available as targets of kinase
inhibition, the more indirect information we have about the
details of the terms of the model. In many models it is possible
as well that the algebraic form of the reaction is
$v=ax_ix_j-bx_kx_h$, this does not affect the modular building of
the network, or our approach, but only the range of proteins
affected.

One way to model kinase inhibition (see \cite{egfmod2}) is to
assume suppression of a target reaction $v$, i.e. $v$ will appear
in the representation of the derivatives of the relevant proteins
concentrations multiplied by a kinase suppression coefficient
$\kappa<1$.

This modeling of kinase inhibitors stems from the assumption that, whatever the current impact of a reaction on the network, a kinase inhibitor targeting that reaction can only slow down its effect on the network in relative terms. Clinical and biological evidences in cancer therapy suggest that the response of patients is often limited by both the low efficacy of drug targeting and by resistance mechanism that cancer cells evolve before and during treatments. Nonetheless, many tumors are dependent (`addicted to') on specific signaling nodes/pathways and even subtle reductions in the total amount of such nodes can impair tumor's homeostasis over time (\cite{kappa1}, \cite{kappa2}). Such evidences are the foundation of combinatorial treatments for cancer and of our assumption that even just a relative, but continuous slowing down of a reaction can have important impact on the network.

Most kinase inhibitors discovered to date are ATP competitive and
present one to three hydrogen bonds to the amino acids located in
the hinge region of the target kinase, thereby mimicking the
hydrogen bonds that are normally formed by the adenine ring of ATP
\cite{zhang}.

Oftentimes kinase inhibitors cross-react, with various degrees
of specificity, either with other kinases among the 518 encoded in the
human genome, or with the abundant nucleotide-binding enzymes that
are present inside cells. The degree of kinase inhibitors
selectivity depends on many factors such as their concentration
and cellular context. Biochemical and cellular assays are
available for the dissection of the specificity range of small
molecules for various kinases, but to date the evaluation of
kinase inhibitor selectivity on an organismal level remains a
significant research challenge \cite{zhang}.

Different models of inhibition of reactions can be easily
implemented in our method, for example forward rate kinase
inhibition and backward rate kinase inhibition would require
$\kappa$ to act only on the first or second term of the reaction
respectively. Since changes in $\kappa$ in general do not affect
to a large extent the activity of any given individual node, in
Section 3 we use $\kappa=0.1$. Such value of $\kappa$ allows for
detectable changes of trajectories, but it may lead in real
systems to unspecific inhibition.

Our discussion up to this point clarifies how the knowledge of a
reaction in the system can be used to build specific building
blocks in selected differential equations of the unknown model.
Next, we write down the heuristic description of a modified
augmented reconstruction algorithm that includes knowledge
of the reactors of a given set of reactions. The details of the algorithm are given in Appendix 2.
\begin{center}
{\bf Modified ASR Algorithm}
\end{center}
Select a collection of potential target reactions
$v_s=a_sx_{i_s}x_{j_s}-b_sx_{k_s}$, $s=1,...,S$. Given the
collection of all time measurements for each node $n$ with
$n=1,...,N$:
\begin{itemize}

\item[{\bf R1}] Set up a representation matrix $Z$ where each
column corresponds to a term of the right hand side of equation
(1) (constant, linear, quadratic and random). Set up a vector
$Y_n$ that corresponds to the left hand side of (1).

\item[{\bf R2}] Select the columns of $Z$ that correspond to the
target reactions involved in the activity of the given node $n$.

\item[{\bf R3}] Perform augmented sparse reconstruction for $Y_n$
using only the columns selected in step $R2$ to force those terms
to have large parameters in the overall representation. Subtract
the contribution of the target terms from the vector $Y_n$.

\item[{\bf R4}] Perform augmented sparse reconstruction for the
modified $Y_n$ using the full representation matrix $Z$. Add back
the the parameters of the target terms found in the previous step
to the corresponding parameters found with the full matrix $Z$.

\item[{\bf R5}] Choose a threshold $T_n$. The reconstructed
network equation for node $n$ will have only linear and quadratic
terms that correspond to parameters larger than $T_n$.

\end{itemize}
The modified ASR algorithm {\bf R1-R5} generalizes the algorithm
in \cite{jtb} in such a way that, for each node, a preliminary
augmented sparse reconstruction is performed only on the terms
that are related to the reactions we selected as potential kinase
targets, if they have an impact on that node. After this
preliminary step, a full augmented sparse reconstruction is
performed with all potential linear and quadratic terms.
We stress that  only the reactors in the targeted reactions need to be known, while the parameters of the reactions are found
by the modified ASR method
automatically.

Note that we output a full model from the algorithm, rather than a
list of directed links to each node. This puts us in the position
of testing our conjecture that the augmented sparse reconstruction
of a network can be homologous to the reference network.

\section{Partial Homology of Networks}

With the algorithm described in the previous section, we can
produce a reconstructed network model of the reference EGF-R
network that is very likely to include the reactions
$v_s=a_sx_{i_s}x_{j_s}-b_sx_{k_s}$, $s=1,...,S$ that we want to
target with kinase inhibitors with nonzero parameters $a_s$ and
$b_s$.

Throughout the remaining sections, it is assumed that we have sets of  $20$
different noisy initial conditions for time courses of nodes of the network, and that we
sample each time course at $11$ points in the interval $[0,11]$,
with time measured in minutes. This interval is acceptable
because, after $t=11$, most nodes relax back to their steady state
and they do not contribute to the understanding of the dynamics.
Noise level for each time course is assumed to be at most $10\%$ of
the maximum value of the points along the time course itself. This
setting gives us a time course microarray with $20 \times 11=220$ data points for each of the $103$ nodes in the reference EGF-R network.

This number of data points in the microarray is very small from the data mining perspective, but it is large from the experimental, {\it in vitro}
point of view. However, it is within the limits of current
experimental practice for reverse phase protein arrays, see for
example the study in  \cite{rppa1} or \cite{rppa2}\footnote{We
mention here as well our work in preparation on adult stem cells,
where the aggregate data set generated by RPPA encompasses about
300 data points per node: Functional Protein Network Activation
Mapping of Adult Mesenchymal Stem Cells Differentiation, B.
McCloud, L. Liotta, E. Petricoin.}.

If a specific experimental setting, or cell type, does not
activate some pathways, it is likely that those pathways will not
be detected, and their contribution to kinase inhibition will not
be measurable. Therefore, initial conditions of protein concentrations, corresponding to  distinct cell states, should ideally be
able to explore as much of the dynamical range of the reference
network as possible. This can be achieved experimentally by varying growth factors concentrations, e.g. EGF, across cell lines of different origins.

At the same time, we only need to detect
homology according to our definition in Section 1, and we argue in
this section that it is possible to generate an homologous
reconstructed network using the modified ASR algorithm even when only very
limited data are available.

In this paper, the time course microarray data used in the modified ASR algorithm are simulated from the reference EGF-R network described in \cite{egfmod1}. The simulated initial time course microarray, and  the modified ASR algorithm, provide us with a candidate reconstructed network that is then compared to the reference network\footnote{ The
values of the time courses are known to be positive, and this is
enforced, for the simulation of the reconstructed network, in the discrete
differential equation solver.}. Reference and reconstructed networks are simulated in the MATLAB environment available at www.mathworks.com.

The choice of meaningful initial
conditions for the time courses is complicated by the fact that the copy numbers of
individual proteins vary enormously, and protein concentration
varies with cell type and cell cycle stage, from less than 20000
molecules per cell for the rarest types to 100 million copies for
the most common ones. In the average mammalian cell some 2000
proteins are considered to be relatively abundant \cite{alberts}, \cite{lodish}.

On the basis of these broad considerations, the initial conditions
for the protein concentrations of the reference network are chosen to be random values uniformly
distributed in the interval $[2000,20000]$. These values are
assumed to be the average number of copies of molecules per cell, and we are assuming that different cell lines might have inherently diverse protein expression patterns. The average number of EGF receptors is taken to be a random value
in the interval $[1000,10000]$ . EGF is selected to be a random
value in the interval $[10^{-8}, 10^{-7}]$ to simulate varying degrees of EGF stimulation. Units are different for EGF as this is a compound
outside the cell and we measure it in gr/ml.

{\bf Remark 2:} Even though this range of concentrations of each
protein is reasonable, the choice of initial conditions for concentrations is still not necessarily biologically meaningful, since the relative
distribution of the initial conditions of the nodes with respect
to each other is randomly selected. Yet we believe that our choice
reflects two basic assumptions that are necessary for the success
of our method, and that are indeed biologically meaningful: strong
variability of time courses, and measurable dynamical changes.

{\bf Remark 3:} Note moreover that the size of the simulated time course microarray  used in the modified ASR reconstruction is very small compared with
the volume of potential initial conditions, so that we are
severely undersampling the space of allowed initial conditions.
Yet, we will see in the following that these small data sets have
predictive power when used to infer the degree of inhibition of
nodes with other initial conditions. Therefore it seems that
homology of networks has some robustness with respect to potential
variations in the experimental setting, so that not every context
relevant to a specific study need to be probed for the application
of our methodology.

{\bf Remark 4:} Our analysis of the simulated EGF-R network model described in \cite{egfmod1} is done assuming that the parameters of the network are constant in the interval of time during which the time course microarray is generated. This assumption seems reasonable because, even in practice, the time courses over which the microarray data are generated are limited to a time scale of days, and in the scenario of our simulations time courses are taken over only 11 minutes. Moreover, in case of translating the modified ASR method to a real world situation where we may seek a therapy for a disease or a pathophysiological state, the time to disease progression and hence to a potential change in the underlying network, is measurable in a time scale of months \cite{timescale}. We conclude that during the time required for applying the modified ASR method to biological samples microarray dataset, the networks' parameters can be considered reasonably constant.

For clinical purposes, not all nodes are of interest. For example
in the reference model of EGF-R, $x_{51}$ (doubly phosphorylated
MEK) and $x_{59}$ (doubly phosphorylated ERK) are particularly
significant targets \cite{egfmod1}. In this paper we are
interested in showing the {\it global} effect of a wide choice of
kinase inhibitors on the ensemble of all network nodes, to
determine a global measure of homology for the reference and
reconstructed networks, with limited data available for the
reconstruction of the homologous network. At the same time
we focus on at least one node (node 31, or Shc) that is particularly amenable to our techniques, and that has biological relevance.

To check the global effect of kinase inhibitors on the network, we
select a set $\mathcal {S}_1$ of $19$ reactions to target with inhibitors, on the basis of
their position along the EGF-R signaling cascade, in order to
comprise both upstream and downstream molecular events that span
the entire signal transduction cascade from top (EGF-R docking
sites) to bottom (MEK or ERK phosphorylarion). The set of reactions $\mathcal {S}_1$ is comprised by 9 pairs of reactions for surface and internalized receptors plus a single degradation reaction\footnote{
The specific reactions that we select in $\mathcal{S}_1$ are:
v19, v66; v20, v67; v23, v70; v27, v74; v29, v76; v41, v83; v45, v87;
v47, v89; v55, v97; and v60. Refer to \cite{egfmod1} for an actual description of these reactions.}.

Although the function of internalization is not the same for all the receptors, it has been demonstrated that EGF-R signals through internalized receptor complexes. A clear correlation between surface and internalized molecules has not been discovered since internalization also causes receptor deactivation. Importantly, \cite{egfmod1} demonstrated that under certain circumstances, such as low EGF concentration, the internalization rate is a critical factor in EGF-R signaling.
We took into account both the surface and internalized reactions since both contribute to the overall signaling and, at least for the EGF-R system, molecules exist already in the clinics which would allow for the selective targeting of surface receptor only (Cetuximab monoclonal antibody). This selective targeting is not feasible for many drug targets that are currently exploited for cancer therapy, but we can still treat in theory the internalized or non-internalized versions of each reaction as if they were two distinct reactions, and hence evaluate their relative impact upon inhibition. Notably, a single node that signals through both internalized and non-internalized counterparts could be inhibited by blockade of both reactions. Similarly, if the internalized version of a node is not affecting the signaling output, the reconstructed system will suggest inhibition of the non-internalized reaction.

We also note that our results do not depend essentially on selecting pairs of internalized and non-internalized reactions. In Figure 4 we summarize results for three other sets of reactions for which there is not extensive matching of internalized and non-internalized versions of the same reaction.  This scenario is the most general one, applicable also to networks other than the EGF-R signaling network, and therefore we will treat internalized and non-internalized version of reactions as if they were distinguishable.

In Figure 1 we can see an instance of the effect of kinase
inhibitors on both reference and reconstructed networks. The top
plot shows the change of time course of node $66$ of the reference
EGF-R network when reactions $v41$ and $v83$ are inhibited with
$\kappa=0.1$. The bottom plot shows the corresponding change of
time course for the $66$th node of the reconstructed network.
The dynamics of reference and reconstructed networks are just
marginally similar for the node in Figure 1. The important point is that we do observe a {\it measurable change} of time courses due to the inhibition of those two reactions. Indeed, we choose to display node $66$ only because its displacements for reference and reconstructed networks are of similar order of magnitudes and amenable to a direct graphic comparison. The similarity of the time courses is, for other nodes, even less pronounced than what we observe in Figure 1 for node 66, or it may happen that the change of time course of a node, due to control, is several order of magnitudes smaller for the reconstructed network. This is expected, since we are
using an incredibly small amount of data to build the
reconstructed network and we cannot expect similarity in the
actual trajectories generated by this coarse approximation.
%FIGURE 1
%
\begin{figure}
\includegraphics[angle= 0,width=0.7\textwidth]{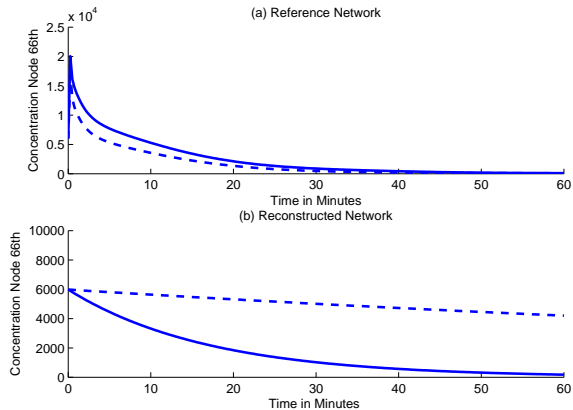}
\caption{\small{An instance of relative concordance of the
magnitude of displacement of trajectories for reference and reconstructed networks due to a specific pair of kinase inhibitors. Dashed curves are the trajectories with
inhibitors on, while solid curves are the trajectories without
inhibitors. The top plot shows the effect of inhibition of reactions
$v41$ and $v83$ on the 66th node of the reference EGF-R network. The bottom plot shows the corresponding trajectories for the
same node $66$ of the reconstructed network. The dynamics of reference and reconstructed networks are just marginally similar for this node. The important point is that we do observe a {\it measurable change} of time courses due to the inhibition of the reactions for both networks.}}
\end{figure}
As an aside, it is not surprising that inhibition of reactions v41 and v83 would naturally lead to the modulation of node 66 (and 35, not shown). In fact, reaction v41, $[(EGF-EGFR^{*})2-GAP-Shc^{*}] + [Grb2-Sos] \leftrightarrow [(EGF-EGFR^{*})2-GAP-Shc^{*}-Grb2-Sos]$, and it's internalized counterpart v83, represent the recruitment of adaptor proteins to the activated receptor. These reaction are directly upstream of node 35 and 66, which are the canonical and internalized versions of the same node (i.e. $(EGF-EGFR^{*})2-GAP-Shc^{*}-Grb2-Sos$). Interestingly, it has been shown that treatment of leukemia cells with a specific Grb2-SH3  (Growth factor Receptor-Bound protein 2 - SH3 domain) inhibitor that disrupts the Grb2-Sos  (Son of Sevenless) complex, has dose-dependent cytotoxic effects, although not comparable with the efficacy of Gleevec in chronic myelogenous leukemia \cite{reff_1}.

In Figure 1, note also that the sign of the change due to control is the
opposite in the two networks. In Appendix 1 we show that this is a common
occurrence with this method. This sign switching may potentially
be lessened by cross-validating parameter estimation for the
different nodes affected by the targeted reactions. We stress that
our goal is not exact trajectory reconstruction, but only an
estimate, using the reconstructed network, of how much the
magnitude of trajectories of the reference network are changed when
kinase inhibitors control is switched on.

To achieve this goal, we propose the following {\bf estimation of the
change of the trajectories due to kinase inhibitors control:}
{\it $\mathbf{(a)}$ We randomly select several initial conditions for concentrations in
the same wide region used to generate the initial time course microarray of the
reference system. $\mathbf{(b)}$ We simulate reference network and
reconstructed  network with each of these initial conditions for a
fixed length of time, long enough for most time courses to relax
to their steady state. $\mathbf{(c)}$ We perform these simulations
with the target kinase inhibitors switched on, and then with the
kinase inhibitors off. $\mathbf{(d)}$ For each node, we compute
the maximum pointwise distance between trajectories with same
initial conditions and with control on and off respectively, over
the whole time interval used in the simulations. $\mathbf{(e)}$
The maximum pointwise distance for each node is divided by the
maximum value of the trajectory of the same node in the reference
network, with control switched off. $\mathbf{(f)}$ Finally, we
measure the {\it median} of the scaled displacement for each node
variable, with respect to the set of initial conditions, as a
statistically significant measure of node displacement due to
control.}

We call the quantity generated by the procedure $\mathbf{(a)-(f)}$
the {\it median scaled maximum pointwise displacement} of
trajectories of a node due to the specific choice of control
kinase inhibitors, sometimes we will refer to this quantity simply
as median scaled displacement.

{\bf Remark 5:} We consider the median since most of the time the
scaled displacement has very similar behavior for most initial
conditions, but there is rarely a small number of initial
conditions that may lead to divergent time courses for some nodes
in the reconstructed system, making the mean too sensitive to
these outliers.

In Figure 2 we plot the magnitude of the median scaled pointwise
displacement of node $31$ with respect to an ordering of
$190$ different kinase inhibitor combinations, where only one or at
most two reactions of the set of 19 reactions $\mathcal {S}_1$ (described in footnote 4) are inhibited at the same time\footnote{ The total of $190$ combinations is the result of taking every combination of 1 out of 19 reactions (19 such combinations) and every distinct combination  of 2 reactions out of 19 ( 171 such combinations).}.
Each kinase inhibitor coefficient $\kappa_s$, $s=1,...,19$ is allowed to be either $1$ (no inhibition) or $0.1$ (high kinase inhibition),
This combinatorial constraint is in line with current experimental
protocols, i.e. to allow only two kinase inhibitor coefficients to
be different from one for each possible combination. In this
scenario an important problem is to find the optimal two kinase
inhibitors to choose from a possibly large collection of
inhibitors. More complex combinatorial scenarios can be envisioned as well, with several kinase inhibitors used at the same time.
Note that generally a kinase inhibitor is considered useful if it
changes the phosphorylation of a target by a significant amount.
Our choice of $\kappa_s$ equal to $0.1$ for individual reactions
increases the chance that we observe relative displacements of the
nodes of the reference network in the order of $10-20\%$.

{\bf Remark 6:} It is implicitly assumed in the model that the
kinase inhibitors are specific at these concentrations.
Notwithstanding the strict 100-fold specificity criteria used
during the screening that companies usually perform for kinase
inhibitors selection process, many inhibitors show various degrees
of off-targets, depending either on their concentration or on the
cell type. Interestingly, some approved drugs (e.g. sunitinib,
dasatinib) had relatively low selectivity but are nevertheless
effective for clinical use. Knowledge of target profiles should
allow careful evaluation as to which drug or drug combination
should be used in a particular situation to better exploit each
drug's full potential \cite{johnson}.
%FIGURE 2
\begin{figure}
\includegraphics[angle= 0,width=0.9\textwidth]{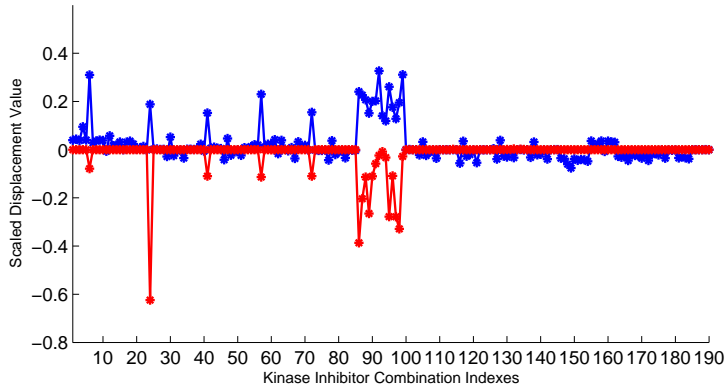}
\caption{\small{In this plot we show a case in which the shape of
the median scaled maximum displacement curve is very similar for
reference and reconstructed networks for high values of
displacement. The blue starred line gives the median scaled
maximum displacements for the 31th node of the reference EGF-R
network. Each point on the horizontal axis corresponds to one
among 190 different pairs of kinase inhibitors combinations. The
red starred line gives the median scaled maximum displacement
curve for the same node in the reconstructed network. Curves are
scaled to have norm equal to one for comparison
purposes. Note the  reversal of sign of the
median displacement curves for large displacement values.}}
\end{figure}

There is a striking concordance of the shape of the two median
scaled displacement curves in Figure 2 for large values of median scaled displacement, even though the magnitude
for each individual kinase inhibitor combination can be vastly
different and indeed the magnitude of the median scaled
displacement for the reconstructed network can be far lower that
the one of the reference network for many nodes, even when there
is very high concordance of the shape of the median scaled
displacement curves. The observed concordance suggests that the median scaled displacement curve of the reconstructed network can be used to infer the corresponding curve for the reference network.

We choose to focus our attention on the 31st node of the reference network as this represents Shc protein (Src homology 2 domain-containing transforming protein 1 or SHC1), an important potential target of effective therapies that acts upstream of the Ras oncogene. More specifically, isoforms p46Shc and p52Shc, once phosphorylated, couple activated receptor tyrosine kinases to Ras and are implicated in the cytoplasmic propagation of mitogenic signals. Thus isoform p46Shc and isoform p52Shc may function as initiators of the Ras signaling cascade in various non-neuronal systems \cite{reff1}. Differently from other isoforms, p66Shc does not mediate Ras activation, but is involved in signal transduction pathways that regulate the cellular response to oxidative stress and life span. Moreover p66Shc acts as a downstream target of the tumor suppressor p53 and is indispensable for the ability of stress-activated p53 to induce elevation of intracellular oxidants, cytochrome c release and apoptosis. In fact, p66Shc competes with other Shc isoforms for binding to the activated receptor complex and skews the signal transduction towards apoptosis. Finally, the expression of p66Shc has been correlated with life span, therefore conferring to such a node a great biological importance \cite{reff2}.
The ERK pathway (Extracellular Signal-Regulated Kinase), which  regulates cell proliferation, is initiated through recruitment to the activated EGFR of a protein complex that contains Shc itself and GRB2 adaptor proteins bound to the exchange factor Sos. Subsequently, Ras is activated by GTP hydrolysis thus leading to ERK phosphorylation and cell proliferation. Ras is one of the most important oncogenes whose deregulation inevitably confers a proliferation advantage to neoplastic cells. In fact, somatic KRAS mutations are found at high rates in leukemias, colon cancer, pancreatic cancer and lung cancer, whereas KRAS gene mutation is predictive of treatment response to anti-EGFR therapy in colorectal cancer patients \cite{reff3}. It is conceivable that the high degree of homology of displacement curves observed for node 31st (Shc) is due to its critical position along the EGF-R signaling and to the major role of the protein complex that is involved in regulating Ras activity. Nonetheless strong homology of median scaled displacement curves has been observed for downstream events (node 58, i.e. $ERK$-$P$-$MEK$-$PP$), or other nodes that are not in direct proximity to EGF-R activation (node 38, i.e. $Shc^*$-$Grb2$-$Sos$). Therefore our method is capable of recapitulating network interactions without specific biases to their location across the network.

In general, the displacement curves tend to agree only partially and only for the
largest displacements values. We also stress the fact that often (but by no means always, see Appendix 1)
the sign of the displacement is different for the reference
network and the reconstructed network, even when we have highly
correlated absolute values of the displacement curves, this is
potentially a problem because the sign of the displacement will
determine whether the control acts as a inhibitor or a enhancer of
the target node.

It is likely that the cause for the sign switching is the
inability of the restricted $l_1$ optimization in step {\bf R2} of
the modified ASR algorithm to detect the proper parameter in the
presence of noise, even when we enforce that such parameter should
be present. Regardless, the displacement curve of the
reconstructed network will be able to identify combinations of
kinase inhibitors that have high impact on the node and in the following we compare only absolute values of median scaled displacement curves.

%figure on Spearman correlation
%FIGURE 3
\begin{figure}
\includegraphics[angle= 0,width=0.9\textwidth]{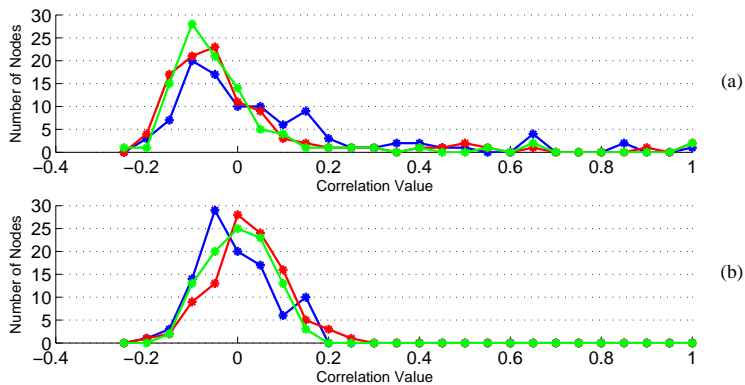}
\caption{\small{In Figure 3(a) we show three histograms of the
Spearman correlation of the absolute value of corresponding median
scaled maximum displacement curves in the reference and reconstructed network.
Each histogram correspond displacement curves generated with one of three different choices of initial time course microarrays as input for the modified ASR algorithm. In Figure 3(b), we show the histograms of the Spearman correlations of the displacement curves of the nodes of the reference network, and randomly permuted versions of the displacement curves of the nodes of the reconstructed networks. All displacement values below $10\%$ of the maximum value of each  displacement curves are set to zero before computing the Spearman correlations.
}}
\end{figure}

For example, we can compare the Spearman's rank correlation \cite{stat} of the absolute value of displacement curves for corresponding nodes in reference and reconstructed networks. A high Spearman's rank correlation  would indicate a very good agreement of the pattern of increase and decrease of large displacements for a given node. In Figure 3(a) we show three histograms of Spearman correlation of corresponding median scaled displacement curves for reference and reconstructed networks. All of them are obtained assuming that the set $\mathcal {S}_1$ of reactions (see footnote 5) is the target of kinase inhibitors, and that the modified ASR algorithm is given as input three different sets of time course microarrays, i.e. three completely different sets of initial conditions and dynamical evolutions of the EGF-R network. To avoid the impact of relatively small displacement values, all displacement values below $10\%$ of the maximum value of each  median scaled displacement curves are set to zero before computing the Spearman correlations. Note the very long right tail of large correlations, indication of excellent agreement of displacement curves for several nodes. Such tail is completely absent when, in Figure 3(b), we compute the histograms of the Spearman correlations of the median scaled displacement curves of the nodes of the reference network, with randomly permuted versions of the median scaled displacement curves of the nodes of the reconstructed networks.

\section{Homologous Control of Nodes}

Our analysis so far gives a sense of the distinctive homology of
reconstructed and reference networks. Our final goal is to use homology
to obtain a nearly optimal kinase inhibitor combination, and we
propose the following strategy:
\begin{center}
{\bf Homologous Control}
\end{center}
\begin{itemize}

\item[{\bf C1}] Given a set of trajectories in a signaling network, and a
set of reactions to be inhibited, use the modified ASR algorithm
in Section 2 to obtain a reconstructed, potentially homologous
network.

\item[{\bf C2}] Consider a large set of kinase inhibitor
combinations that satisfy some give constraint. For each kinase
inhibitor combination generate time courses of the reconstructed
network for a variety of biologically meaningful initial
conditions.

\item[{\bf C3}] Generate the median scaled displacement curve for
a target node protein. Identify the position of the few largest
values of the median scaled displacement curve for the
reconstructed network. The corresponding kinase inhibitor
combinations are candidates for nearly optimal
suppression/enhancement of the target node in the reference
network.

\end{itemize}
Several possible choices of constraints could be enforced in step
{\bf C2} of this algorithm. For example, in personalized therapies
we could ask for the combination of kinase inhibitors with a
minimum total norm of the corresponding kinase inhibitor
coefficients $\kappa$, to reduce total amount of inhibitors and therefore to avoid toxicity and loss of specificity.

In the following, we continue to explore the experimental protocol
used to generate Figures 2 and 3 in which only two reactions at the
time are inhibited. Recall that since we have $19$ possible kinase
inhibitor targets, there are a total of $190$ distinct pairs and
singlets of kinase inhibitor therapies.

In our simulations there is remarkable agreement of the location
of large peaks of the median displacement curves for reference and
reconstructed networks, so that it seems possible to use the
largest median displacement values of the reconstructed network to
predict the likely location of near-optimal combinatorial kinase
inhibitions.

We cannot expect full overlapping of locations of large peaks and
therefore we suggest the following  definition of {\bf
near-optimality of kinase inhibitor combinations}: {\it We assume
that we found near-optimal kinase inhibitor combinations if the
locations of the top three maxima in the median scaled displacement curve
of a node of the reconstructed network overlaps with the location
of large median scaled displacement values of the corresponding
node in the reference network. A large displacement is defined
here as a value that is a large percentage, say $80\%$, of the
mean of the largest 3 displacement values of the given node in the
reference network.}

{\bf Remark 7:} Note that we use, as a benchmark of success, the
mean of the largest three median displacement values, rather than the
absolute maximum displacement. We decided for this criterium since
in principle the absolute value may be so large compared to the
other displacements values that there may not be other significant
kinase inhibitor combinations with displacement values close to
the maximum.

Since the scaled displacement curves of many nodes of the
reconstructed network are not carrying any useful information on
the corresponding nodes of the reference network, we need to find
a way to filter the most useful nodes of the reconstructed
network.

Our understanding is that the larger the displacements of a node in the reconstructed network, the more likely the chance that they convey some useful information about the corresponding displacements of the node of the reference network. This is the case because large displacements probably indicate an increased sensitivity of the reconstructed network to specific kinase inhibitors. The following procedure defines quantitatively a {\bf notion of large displacements:} {\it we take the median nonzero displacement value for each node in the reconstructed network, and then the mean of these medians across all the nodes in the network. We  threshold to zero all displacements that are below this value, so that we retain only displacement values that are large relatively to the overall activity of the network.}

After removing small displacements with the previous procedure, we filter the nodes in the network by retaining only nodes that have the mean of the remaining nonzero scaled displacements above increasing large threshold values. Note that these threshold values are dimensionless, since they relate to the scaled displacements that are by definition dimensionless.

This filtering of nodes is very stringent, in the sense that only a few nodes are left by this process, and yet it is quite effective at identifying those displacement curves in the reconstructed network that partially correlate to the corresponding displacement curves of the reference network. In particular, in Figure 4(b), the dashed curves shows the number of nodes of
the reconstructed network that have mean of the large nonzero median
displacement values above the threshold specified on the
horizontal axis. Again, we show our results for three different sets of time course microarrays given as input into the modified ASR algorithm.
%FIGURE 4
%
\begin{figure}
\includegraphics[angle= 0,width=0.8\textwidth]{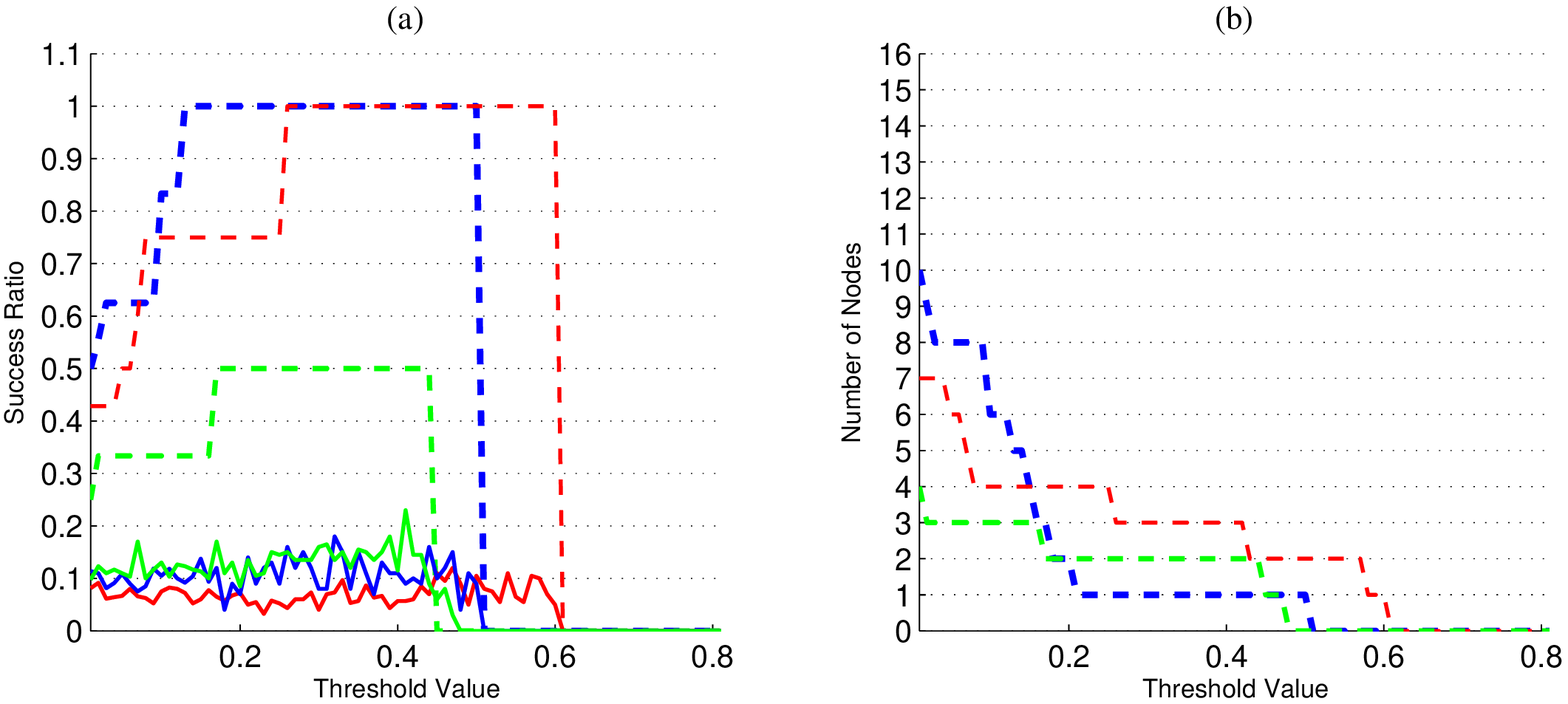}
\includegraphics[angle= 0,width=0.8\textwidth]{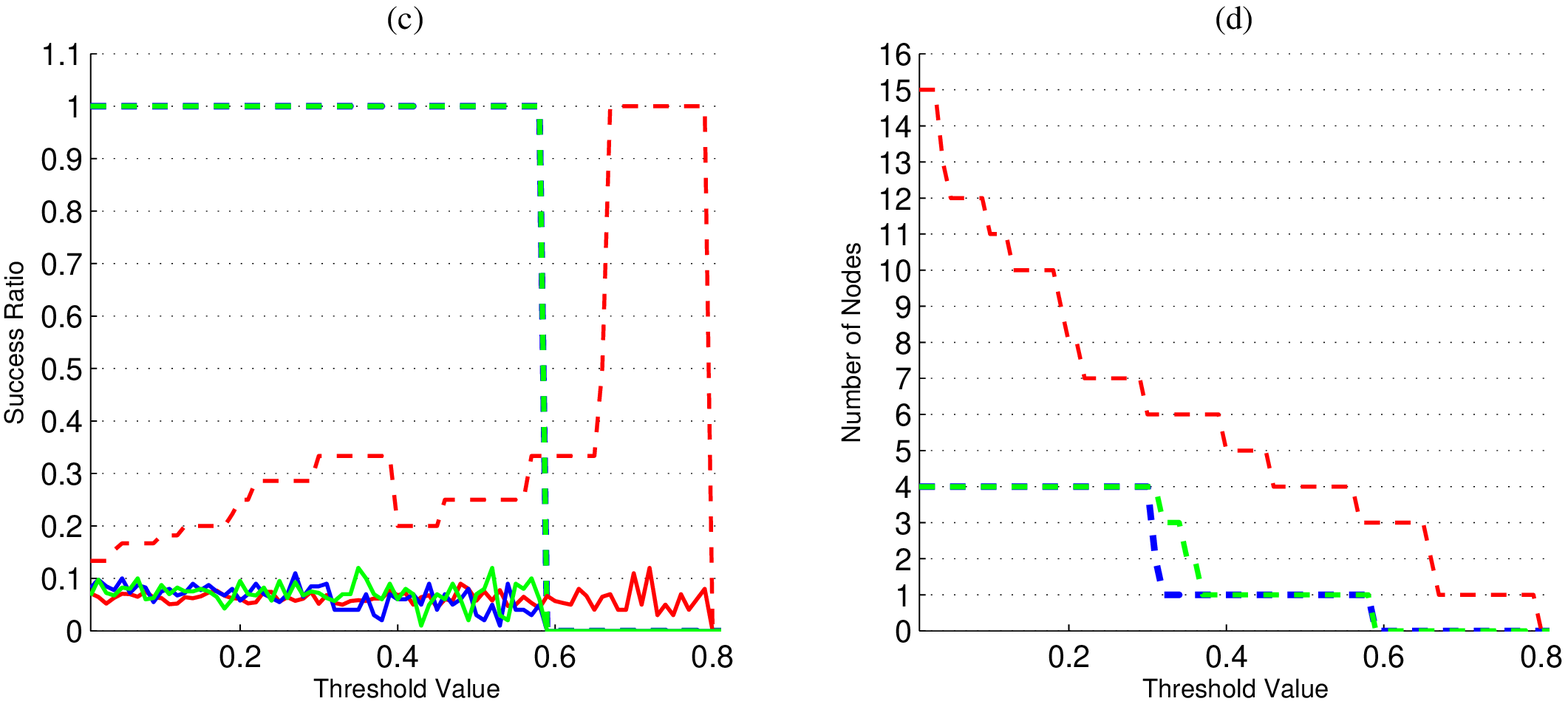}
\includegraphics[angle= 0,width=0.8\textwidth]{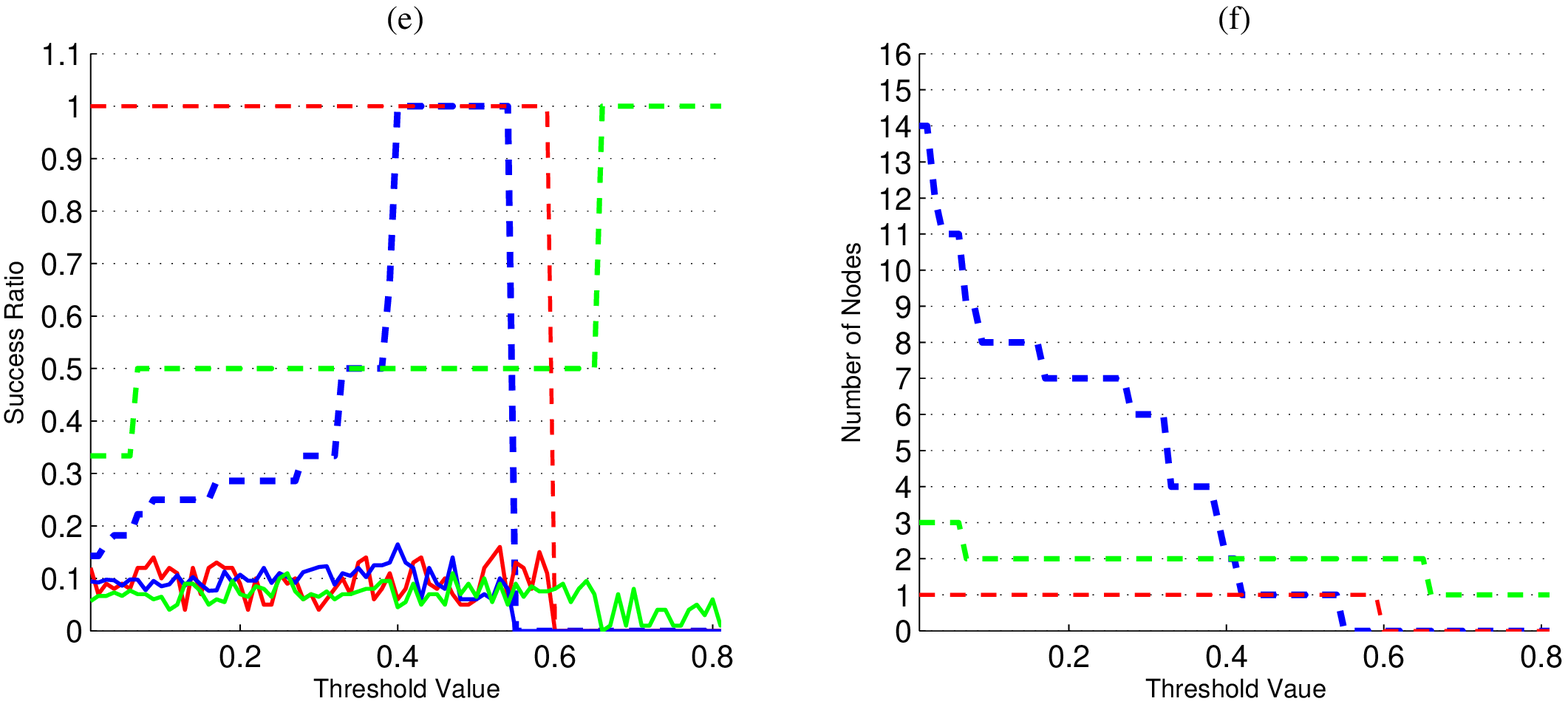}
\includegraphics[angle= 0,width=0.8\textwidth]{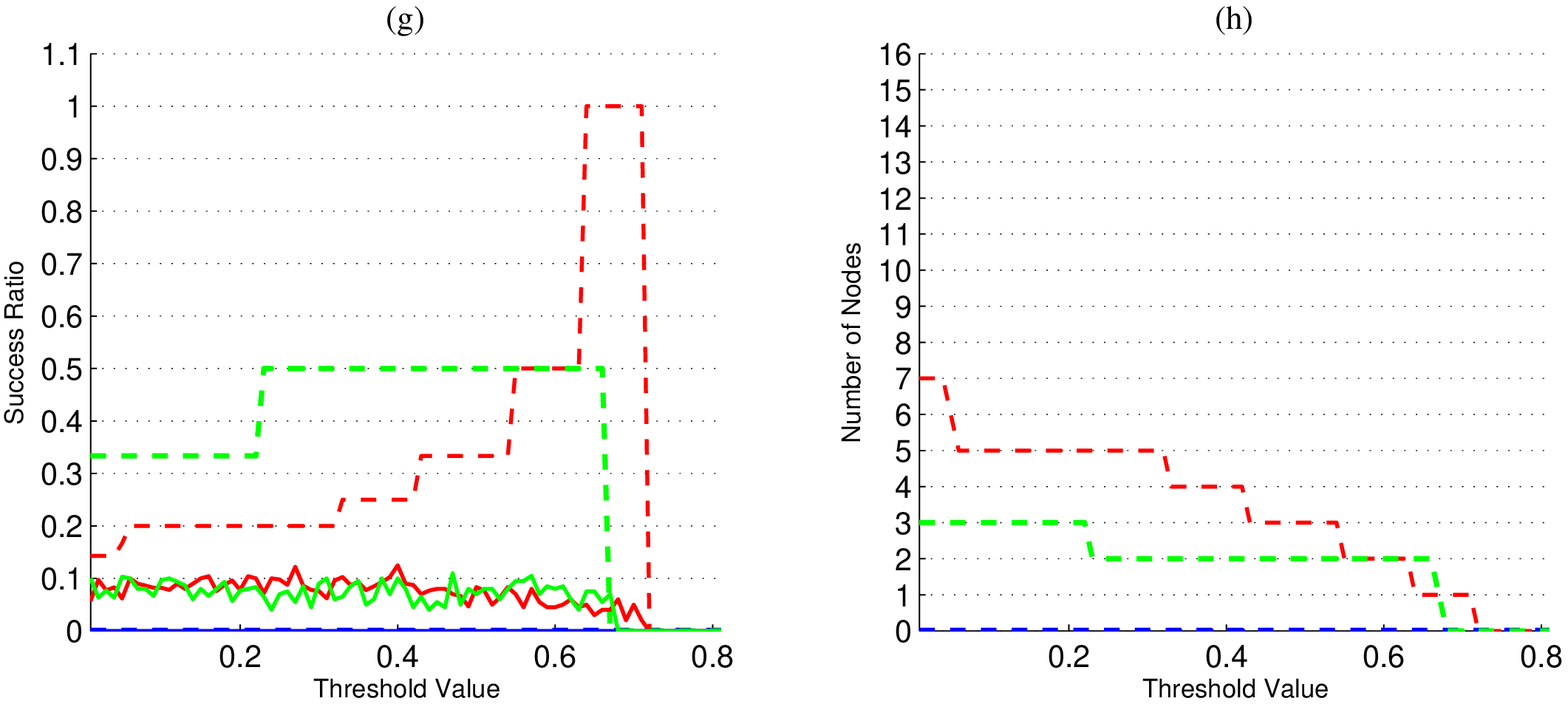}
\caption{\small{Plots (a) and (b) refer to the set of 19 reactions $\mathcal{S}_1$.
In plot (b) the three dashed curves shows the number of nodes of the reconstructed
network that have mean of large median displacement values
above the threshold specified on the horizontal axis.
In plot (a), the three dashed curves shows the ratios of
nodes for which we identify near-optimal kinase inhibitor
combinations, as a function of the same
threshold. The solid curves shows the average ratio of
nodes for which we identify near-optimal kinase inhibitor
combinations when the sequence of median scaled displacements for
the reconstructed network has been randomly permutated (average
computed over $100$ permutations). Each color corresponds to the analysis of one of three different choices of initial time course microarrays as input for the modified ASR algorithm. Plots (c), (e), (g), and (d), (f), (h) repeat the same analysis as in plots (a) and (b) respectively, for three more sets of kinase inhibitors (see footnote 6).}}
\end{figure}

The dashed curve in  Figure 4(a) shows the ratio of
nodes for which we identify near-optimal kinase inhibitor
combinations (as defined above), as a function of the same
threshold as in figure 4(b). We reach at least a $0.5$ success ratio, meaning that, given sufficiently high threshold, for
$50\%$ of the nodes that are not filtered out by the threshold, we can identify
near-optimal kinase inhibitor combinations. The catch is of course
that only a few nodes display such marked homology in
the reconstructed median scaled displacement curve. The $0.5$ ratio mark is reached when the threshold leave respectively 10, 6, and 2 nodes for the three time course microarray inputs to the modified ASR algorithm. On the other hand, it is striking that we can eventually reach $100\%$ success ratio (respectively for 6 and 3 nodes) for two of the microarray inputs .

The solid curves in the same plot 4(a) shows the average
percentage of nodes for which we identify near-optimal kinase
inhibitor combinations when the sequence of median scaled
displacements for the reconstructed network has been randomly
permutated (average computed over $100$ permutations). Near-optimality for the random scrambling of the data is much
lower, especially for the nodes left by higher threshold values. The mean of the ratio of nodes
for which we find near-optimal combinations in the case of random
permutation of the reconstructed displacement curves in Figure
4(a) is only around $0.1$ (computed across all threshold values and all three time course microarrays inputs). For nodes selected with high values of threshold, our method is at least about 5 times more accurate than a random
selection of kinase inhibitors in finding near-optimal
combinations, and up to 10 times more accurate for some selected nodes with two microarray inputs.

To put things in perspective, with the randomly permuted scaled
displacements, on average it would be necessary to select randomly between 18 and 30 kinase inhibitor combinations (depending on the microarray input instance) to make sure that we have at least a $0.5$ success ratio, reachable with only 3 kinase inhibitor combinations by the homologous control scheme. This
represents a large potential saving, due to the data-based
narrowing of combinatorial possibilities, in the search for
appropriate kinase inhibitor therapies.

Figure 4(c,d), 4(e,f), 4(g,h) show the results of the same analysis
performed in Figure 4(a,b), for three other selections of 19 kinase
inhibitors \footnote{The reactions selected for inhibition
corresponding to Figure 4(a,b), 4(c,d), 4(e,f), 4(g,h) are respectively:
({\bf$\mathcal {S}_1$}) v19, v20, v23, v27, v29, v41, v45,  v47, v55, v60, v66,
v67, v70, v74, v76, v83, v87, v89, v97; ({\bf $\mathcal {S}_2$}) v1, v2, v3,  v10,
v16, v28, v29, v36, v37, v45, v46, v47, v48, v64, v75, v94, v95,
v126, v130; ({\bf $\mathcal {S}_3$}) v1, v2, v12, v23, v27, v33, v34, v36, v45,
v53, v56, v74, v76, v78, v89, v97, v111, v129, v130; ({\bf $\mathcal {S}_4$}) v1,
v3, v11, v18, v22, v26, v29, v33, v39, v42, v44, v52, v65, v67,
v75, v88, v96, v128, v148. Refer to \cite{egfmod1} for an actual
description of the reactions.}, and several distinct initial microarrays. The main criterion underlying the
choice of reactions has been the availability, for most of the
selected reactions, of inhibitory drugs that might be used to
block the downstream signaling. Several example exist that have
either passed FDA approval or are currently in clinical trials,
like monoclonal antibodies or small molecules inhibitors that
target EGF-R (Cetuximab and Gefitinib, respectively) and small
molecule inhibitors of the Raf/MEK/ERK axis (Sorafenib). The
number of reactions that satisfy this experimental plausibility
criterium is relatively small in the cited EGF-R network, and
there is a certain degree of overlapping among these sets.

Note that the success curve
is not monotone for one instance (red dashed line) in Figure 4(c), indicating that some strongly homologous
nodes are actually removed as the threshold value increases. This raises the
hope that even better success rates could be achieved with more
sophisticated threshold processes. In particular, it may be beneficial in some cases to have an adaptive threshold that is a variable multiple of the mean of medians used to pre-condition the displacement curves. This may allow for a larger number of nodes to be retained.

The success rate of homologous control is, for high enough threshold value, at least $50\%$ for reactions
analyzed in Figure 4(c,d) and Figure 4(e,f). Figure 4(g,h) is, instead, significantly
different, in that, for one instance of microarray (blue dashed curves) no displacement curve of the reconstructed network passes the filtering process. A careful analysis of the different selections of kinase
inhibitors shows that, while their distribution in the network is
very similar, the actual parameters of the reactions selected for
inhibition are significantly different for Figure 4(g,h). Specifically, the
number of large (bigger than one) forward and backward kinetic
rates is about a third of the corresponding number of large
kinetic rates in the other three selections of 19 kinase
inhibitors. This observation may justify the failure of our method for one instance of initial microarray and it is an important point that both shows the limits
of our technique, and suggests that, for the method to be
eventually applied in practice, care should be made to select,
when possible, kinase inhibitors that act on relatively fast
reactions. Most nodes that show sizable displacement curves are
likely to be close to these fast reactions, when targeted by
kinase inhibitors.

{\bf Remark 8:} Another point that needs clarification is the
extent to which the modified ASR method is essential to the
success of the identification of near-optimal combinations. In
other words, is it necessary to know the nodes that are involved
in each of the 19 reactions that we target for inhibition? The
answer is affirmative: without knowledge of {\it the presence} of
these reactions, many of the relevant parameters are overlooked by
the standard ASR algorithm, in the sense that either they are not
found, or their value is underestimated. We find ourself in a
scenario very similar to one of the instances of Figure 4(g,h), where very few
nodes in the reconstructed network have any response to the kinase
inhibitors (not shown). In general, we expect the performance of the method to improve with larger sets of kinase inhibitors and to worsen
with smaller sets. However, large sets of kinase inhibitors are
likely to arise in the experimental setting and this is the
scenario where the method should work best.

\section{ Discussion and Conclusion}

Our method for homologous control is an attempt to develop a
signal processing approach to network dynamics and it has the
potential of greatly reducing the experimental load necessary to
find near-optimal combinations of kinase inhibitors for a list of
potential target reactions. Its strength is in the ability to work
with very limited, noisy data and with networks that have a large
number of nodes, comparable with realistic time course microarray
data. In this last section we would like to point out several
broad areas for further development that are intimately related to
the limitations of current experimental protocols for measuring
node activity of signaling networks.

Our approach is dependent on the choice of a region $\mathcal R$
where we select the initial conditions, and on the duration $T$ of
the time series, so that we could say that the notion of
homologous systems and signal processing of networks is transient
based, i.e. it depends on the choice of the cylinder $\mathcal R
\times[0,\,\,T]$. This raises some issue on the stability of the
homology, if we run the reference system for a time $T$, how long
should we run the reconstructed system?

One characteristic of the reconstructed network is that  its
dynamics displays slower changes when compared to the reference
network, probably because the parameters of each term in the
equations are not as large as the true parameters, so that the
rate of change of nodes will generally be different for the
reference network and the reconstructed one. Nevertheless, the
type of curves that we observe are usually transients with
eventual relaxation, so that $T$ can be chosen for both networks
as the time such that either the trajectories of the networks have
relaxed to their steady state, or they show consistent divergence.

A possible strategy to improve parameter estimation for large
networks is to run the reconstruction algorithm several times with
different choices of the random terms; collect for each node the
terms that display significant activity for at least one
repetition of the reconstruction algorithm; repeat one last time
the reconstruction for each node, only using the nodes previously
selected as significant. This bootstrap version of the
reconstruction algorithm may improve homology with the reference
system.

An important question is to determine how infrequently we can
sample the trajectories of a network and infer a well behaved
reconstructed network that is homologous to the reference network.
In some sense, we need a sampling theory of networks; note that
the sampling is done for the trajectories, and it is a signal
processing operation, but the notion of well behaved system is
essential a dynamical one.

To gain a sense of the difficulty of this endeavor, consider that
the trajectories' sampling rate used in this paper clearly do not
allow for high true positive rates and low false positive rates of
identification of parameters in the reconstructed network.
Significant noise is observed in the network parameters'
estimation, even in the parameters of the reactions selected for
kinase inhibition.

Indeed, the actual trajectories generated by the reconstructed
network do not need to show any strong resemblance to those of the
reference network. The very notion of homologous networks is
designed to be useful exactly when we undersample the network
trajectories so severely that we do not hope for a proper network
reconstruction. As we showed in this paper, as little as 220 data
points per node are sufficient to obtain partial homology,
possibly because the sensitivity of the initial dynamics of the
trajectories to kinase control may be predictive of the
sensitivity over the full transient leading to relaxation to the
steady state. The theoretical determination of the relation
between number of samples per node and network homology will
require extensive study and comparison of several biologically
meaningful models of pathways.

It is also crucial to understand the performance of our method
when dealing with incomplete networks where only a portion of the
nodes is measured. We tested, for example, the nearly-optimal
kinase prediction algorithm {\bf C1-C3} on a module of the EGF-R
model comprising only 16 variables, with only two reactions
selected for kinase inhibition, and we were indeed able to observe
homology for several nodes even though many nodes were subject to
large feedbacks from unmeasured nodes\footnote{In this simulation (data not shown)
we used 550 data points, and initial condition for each node were kept
very low for this simulation, while EGF was very large.}

It is yet to be seen whether a random choice of a subset of nodes
belonging to a pathway are sufficient to achieve homologous
control. Of course, for our method to make sense, at least all
nodes involved in reactions to be inhibited and the target node
must be measured. We need to perform a detailed study in which we
identify the minimum number of variables (and their distribution)
that need to be measured to achieve homologous control.

The final goal of our approach is the experimental validation of homologous control over a broad range of signaling networks in defined biological contexts such as cell proliferation, survival and differentiation. The availability of hundreds chemical compounds with known specific inhibitory activity allows to test the efficacy of the predicted near-optimal kinase inhibitor combinations into cell line models. In particular, it is possible to exploit cancer cell lines obtained from diverse types of tumor and to train the modified ASR method with reverse-phase protein microarray data containing time-courses for each cell line under serum addition or hypoxic stimuli. The molecular dynamics driven by such conditions  would allow a validation of the {\it in silico} reconstructed network in a specific and biologically meaningful context. The combinations of inhibitors that are found by our method in this scenario could then be tested {\it in vitro} (and eventually {\it in vivo} in immunodeficient mice)  to understand their ability to affect proliferation and survival mechanisms of tumor's cells. Such an approach has a great potential in the identification of novel therapeutic strategies for cancer. We stress moreover that there is no technical reason to restrict homologous control strategies to the analysis of the EGF-R signaling network, or even to protein signaling networks. A more general choice of terms for the model in equation (1) would allow our method to be tested for multiscale heterogeneous systems, where genomic, proteomic and metabolic compounds are related in a single network.

\section*{Acknowledgments}

The authors acknowledge the support of the College of Science at George Mason University and the Istituto Superiore di Sanit\`{a}. They also would like to thank Daniele C. Struppa for many useful discussions and the anonymous referees for their constructive remarks.

\section*{Appendix 1: Sign Concordance and Sign Switching}

One difficulty summarizing the qualitative properties of
displacement curves described in Section 3 is the tremendous variability of the individual nodes of the network. However, it is clear that sign switching and
sign concordance of the displacement curves for high displacement
values are both common, and not due to random effects.

For example, in Figure 5(a) we show the number of kinase inhibitor
combinations, out of the total 190 combinations used in the protocol of Figures 2 and 3, that display opposite sign, as
a function of a threshold on displacement curves of both reference
and reconstructed networks. For each node, the threshold sets to
zero all displacements that are below the percentage of the
largest displacement value denoted in the ordinate axis. Red
curves are for the 5 nodes that have the highest number of
combinations with sign switching. Blue curves are generated in a
similar fashion, but after the median scaled displacement values
have been permuted within the displacement curve of each node.

Note the very distinct behavior at the $20\%$ threshold level
between true and randomized curves in Figure 5(a). Instead, Figure
6(a) shows, for each node of the reconstructed network, the ratio
of nonzero median displacement values that display sign switching
with respect to the reference network, in the case of a $20\%$
threshold. Many nodes, at this threshold level, exhibit large
percentages of sign switching.

%FIGURE 5
\begin{figure}
\includegraphics[angle= 0,width=1\textwidth]{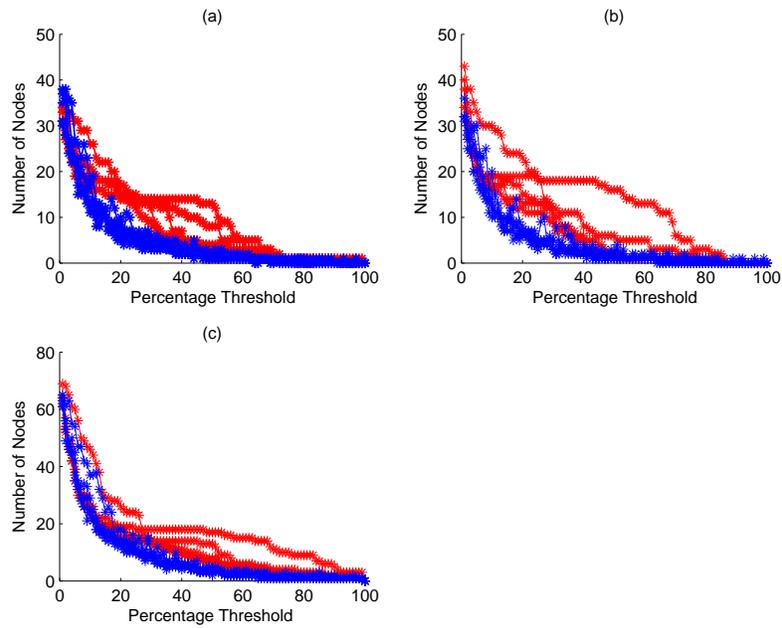}
\caption{\small{Red starred curves in Figures 5(a), 5(b), 5(c) shows
the number of kinase inhibitor combinations of the 5 nodes of
reference and reconstructed networks that display respectively the
largest: 5(a) sign switching; 5(b) sign concordance; 5(c)
concordance of nonzero median displacement values. Blue starred
curves are generated in a similar fashion, but after the median
scaled displacement values of the reconstructed network have been
permuted within the displacement curve of each node. Curves are
plotted as functions of a threshold on displacement curves of both
reference and reconstructed networks. For each node, the threshold
sets to zero all displacements that are below the percentage (of
the largest displacement value) denoted in the ordinate axis.}}
\end{figure}
%FIGURE 6
\begin{figure}
\includegraphics[angle= 0,width=1\textwidth]{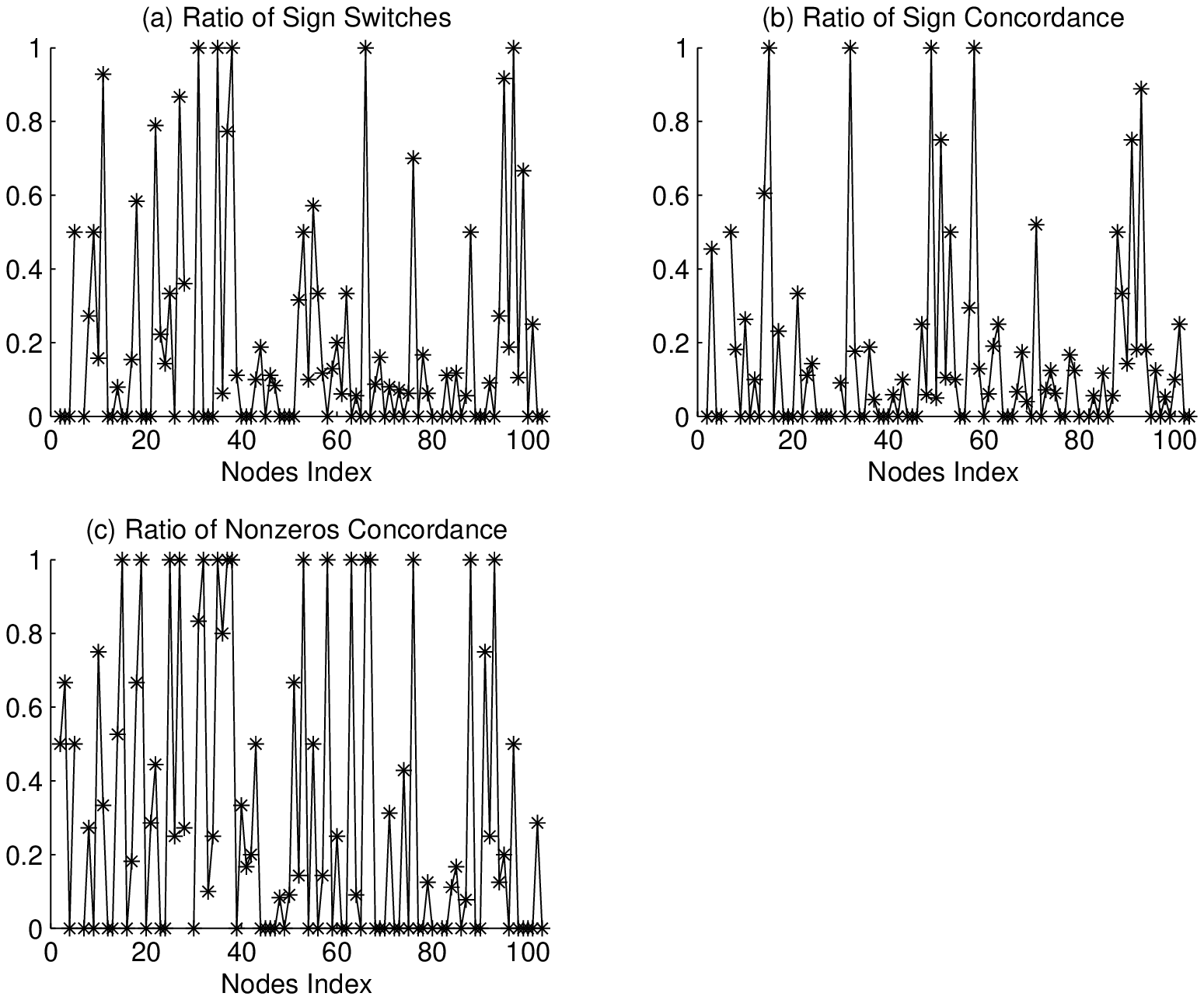}
\caption{\small{  Figures 6(a), 6(b), 6(c) show, for each node of
the reconstructed network, the ratio of nonzero median
displacement values that display respectively: 6(a) sign switching
at the $20\%$ threshold level; 6(b) sign concordance at the $20\%$
threshold level; and 6(c) concordance of nonzero displacement
values at the $40\%$ threshold level. For each node, the threshold
sets to zero all displacements that are below the given percentage
of the largest displacement value.}}
\end{figure}

{\bf Remark A1:} We stress that this sign switching is by no means
consistently observed for all nodes, or for specific groups of
kinase inhibitor combinations. For example, the kinase inhibitor
combinations corresponding to indexing from 85 to 100, shown in
Figures 2 to have vigorous control on at least a node of the
network, have roughly the same number of nodes that display
uniform sign switch and uniform sign concordance for large
displacement values. If we study sign concordance rather than sign
switching, a very similar qualitative behavior as in Figures 5(a)
and 6(a) is observed, as can be inferred from Figures 5(b) and
6(b).

In Figures 5(c) and 6(c) we follow essentially the same procedure
as the one that we used to explore sign switching, but focusing on
the concordance of nonzero displacements. In Figure 5(c) we show
the number of kinase inhibitor combinations, out of the total 190,
that display concordance of nonzero displacement values, as a
function of the same threshold on median displacement curves of
both reference and reconstructed networks used in Figures 5(a) and
5(b). Red curves are for the 5 nodes that have the highest number
of combinations with corresponding nonzero median displacement
values. Blue curves are generated from permuted median
displacement curves.

In Figure 5(c) we see distinct behavior between real and randomized
data especially around the $40\%$ threshold level. The fact that
this percentage is higher than the one observed for the analysis
of signs shows that sign concordance and sign switching are more
statistically significant than nonzero concordance for lower
threshold. Figure 6(c) shows, for each node, the ratio of nonzero
displacement values for the reconstructed network that is matched
by nonzero displacement values for the corresponding node of the
reference network, at the $40\%$ threshold level. Many nodes, at
this threshold level, exhibit almost complete overlapping of the
nonzero displacement values of reconstructed network with (some
of) the nonzero displacement values of reference network.

\section*{Appendix 2: Recursive Augmented Sparse Reconstruction with selected target reactions}

In this appendix we give details of the recursive augmented sparse
reconstruction algorithm to be used in the presence of target
reactions. Suppose we are given $N$ node variables from a network
and that for each variable it is possible to generate $R$
trajectories $X_{n,r}$ $r=1,...,R$ with different initial
conditions, uniformly sampled at $L$ points. We build now the left
hand side of (1) and the individual terms in the right hand side.

Call $\bar X_{n,r}$ the vector $X_{n,r}(t)-X_{n,r}(t_0)$ where $t$
takes all $L$ sampled values. For a given vector $g(t)$,
$t=t_0,...,t_L$, let $I(g)$ be the vector whose $l$-th component
is the sum $\sum_{i=0}^l g(t_i)$.

Write $Y_n=[\bar X_{n,1},...,$$\bar X_{n,R}]$,
$G_n=[I(X_{n,1}),...,$ $I(X_{n,R})]$, $n=1,...,N$, $G_{ij}=[I(
X_{i,1}X_{j,1}),...,$$I(X_{i,R}X_{j,R})]$.
Finally, let $J$ denote
the unit vector with same length as $Y_n$.

Select a collection of potential target reactions
$v_s=a_sx_{i_s}x_{j_s}-b_sx_{k_s}$, $s=1,...,S$. The basic process
to identify the links among the nodes is the following. For each
node $n$ with $n=1,...,N$:
\begin{itemize}

\item[{\bf R1}] Choose an attenuation coefficient $\beta_q$ for
the quadratic terms $G_{ij}$. Let $n_g$, $g=1,..,G$, be discrete
random vectors normally distributed scaled to have norm 1. Denote
by $|\,|$ the $2$-norm of a vector and let $\hat G_l$ be the
matrix whose columns are all the vectors $\frac{G_i}{|G_i|}$,
$\hat G_{q}$ be the matrix whose columns are all possible vectors
$\frac{G_{ij}}{|G_{ij}|}$. Let $N_G$ be the matrix whose columns
are the random vectors $n_g$ scaled to have norm $1$. Choose $G$
large enough to have the matrix $Z=[J \hat G_l, \beta_q \hat G_q,
N_G]$ with small condition number (say less that $10^2$).

\item[{\bf R2}] Set a temporary representation matrix $M$, for
each $s=1,...,S$, if the node $n$ belong to the set
$\{i_s,j_s,k_s\}$, add the vectors $\frac{G_{k_s}}{|G_{k_s}|}$ and
$\frac{G_{i_sj_s}}{|G_{i_sj_s}|}$ as columns to the matrix $M$.

\item[{\bf R3}] Let $Z_M=[M\,\,N_G]$. Find the minimal $l_1$
solution to the underdetermined system $Y_n=Z_M\alpha_M$. Let
$\alpha_M$ be the restriction of $\alpha$ to the columns of $M$,
set $Y_n=Y_n-M\alpha_M.$

\item[{\bf R4}] Find the minimal $l_1$ solution to the
underdetermined system $Y_n=Z\alpha$. If in part {\bf R2} we
generated a nonzero matrix $M$, then add to the components of
$\alpha$ associated to the columns of $M$ the corresponding
components of  $\alpha_M$.

\item[{\bf R5}] Choose a threshold $T_n$ and let $\alpha_{T_n}$ be
the coefficients in $\alpha$ larger than $T_n$. The reconstructed
network equation for $x_n$ will have only linear and quadratic
terms that correspond to coefficients in $\alpha_{T_n}$, and their
coefficients will be the coefficients of $\alpha_{T_n}$ divided by
the norm of the corresponding $|G_{i}|$, if a linear term, and
$|G_{ij}|$ if a quadratic term.

In \cite{jtb} we showed that there is considerable flexibility in
the choice of the number $G$ of random terms and in the choice of
the attenuation coefficient $\beta_q$; in this work we use
$G=1500$ and $\beta_q=0.8$. The threshold $T_n$ that selects the
parameters to be used in the reconstructed network is taken to be
a very low $2\%$ of the maximum magnitude parameter for the
corresponding node. This choice makes sure that most inferred node
directed links among nodes are retained.

\end{itemize}

\end{document}